\documentclass[journal=jpclcd,manuscript=letter]{achemso}
\usepackage{braket}
\usepackage{xcolor}
\usepackage{graphicx}
\usepackage{lineno}
\usepackage{hyperref}
\usepackage{siunitx}
\usepackage{textcomp}
\usepackage{caption}
\usepackage[version=3]{mhchem} % Formula subscripts using \ce{}

\author{Nicola J. Fairbairn}
\affiliation[UEC]
{Department of Engineering Science, Graduate School of Informatics and Engineering, The University of Electro-Communications, 1-5-1 Chofugaoka, Chofu, Tokyo 182-8585, Japan}
\author{Koya Onoda}
\affiliation[UEC]
{Department of Engineering Science, Graduate School of Informatics and Engineering, The University of Electro-Communications, 1-5-1 Chofugaoka, Chofu, Tokyo 182-8585, Japan}
\author{Ozora Iso}
\affiliation[UEC]
{Department of Engineering Science, Graduate School of Informatics and Engineering, The University of Electro-Communications, 1-5-1 Chofugaoka, Chofu, Tokyo 182-8585, Japan}
\alsoaffiliation[UofG]
{School of Chemistry, University of Glasgow, Glasgow, G12 8QQ, United Kingdom}
\author{Masahiro Yabuno}
\author{Hirotaka Terai}
\author{Shigehito Miki}
\affiliation[NICT]
{Advanced ICT Research Institute, National Institute of Information and Communications Technology, 588-2 Iwaoka, Nishi-ku, Kobe, Hyogo, 651-2492, Japan}
\author{Yuji Kagotani}
\affiliation[QST]
{Institute for Quantum Life Science, National Institutes for Quantum Science and Technology, 4-9-1 Anagawa, Inage-ku, Chiba,  263-8555, Japan}
\author{Motoyasu Adachi}
\affiliation[QST]
{Institute for Quantum Life Science, National Institutes for Quantum Science and Technology, 4-9-1 Anagawa, Inage-ku, Chiba, 263-8555, Japan}
\author{Gordon J. Hedley}
\affiliation[UofG]
{School of Chemistry, University of Glasgow, Glasgow, G12 8QQ, United Kingdom}
\author{Ryosuke Shimizu}
\email{r-simizu@uec.ac.jp}
\affiliation[UEC]
{Department of Engineering Science, Graduate School of Informatics and Engineering, The University of Electro-Communications, 1-5-1 Chofugaoka, Chofu, Tokyo 182-8585, Japan}
\alsoaffiliation[UECAdvancedcenter]
{Institute for Advanced Science, The University of Electro-Communications, 1-5-1 Chofugaoka, Chofu, Tokyo 182-8585, Japan}

\title{Development of time-resolved fluorescence spectroscopy using heralded emission spectroscopy for organic light-emitting materials}

\begin{tocentry}
\centering
  \includegraphics[width=\linewidth,height=1.6in,keepaspectratio]{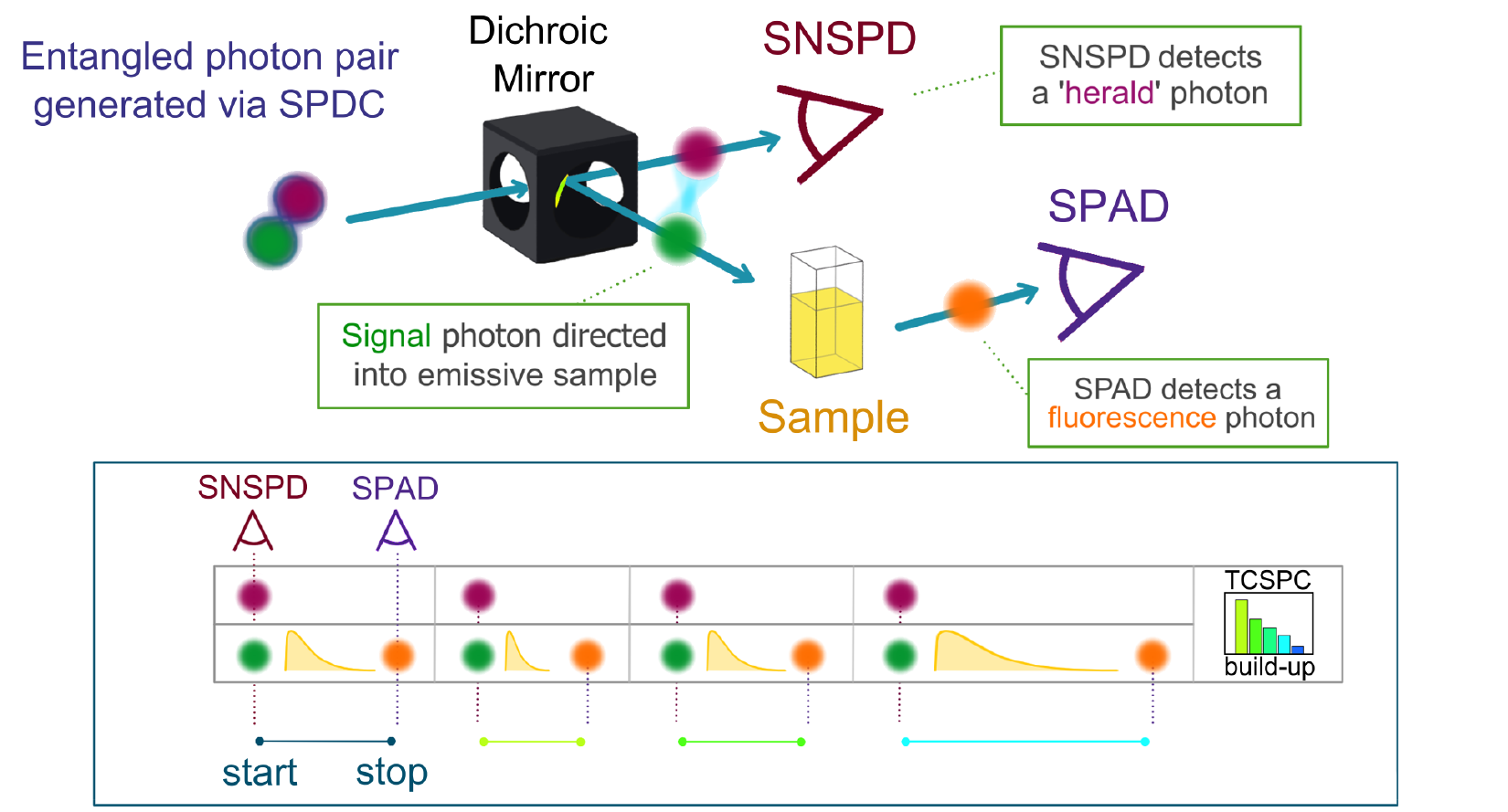}

\end{tocentry}

\begin{document}
\newpage

\begin{abstract}
Organic light-emitting materials are widely used in optoelectronics and bioimaging, but sometimes conventional fluorescence spectroscopy techniques can be limited by scattering, background fluorescence, photon pile-up and photobleaching. We report a quantum-spectroscopic approach using frequency-entangled, nondegenerate photon pairs generated via spontaneous parametric down-conversion with heralded detection, to perform photoluminescence lifetime measurements at low excitation intensities. We demonstrate this approach across three organic light-emitting material classes: the fluorescent dye Rhodamine 6G, an organic semiconducting conjugated polymer poly[2-methoxy-5-(2-ethylhexyloxy)-1,4-phenylenevinylene] (MEH-PPV), and a fluorescent protein mOrange. The intrinsic timing correlation between signal and idler photons enables coincidence gating that suppresses uncorrelated background counts. Across all materials and integration times of 500, 1000 and 3600 seconds, heralded detection yields signal-to-noise ratios 1.6-3.0 times higher than classical pulsed-laser excitation attenuated to match heralded photon counts in the fitting range. For MEH-PPV, this roughly halves the noise amplitude. Our measured lifetimes closely match long-integration classical pulsed-laser reference values.

\end{abstract}
%%%%%%%%%%%%%%%%%%%%%%%%%%%%%%%%%%%%%%%%%%%%%%%%%%%%%%%%%%%%%%%%%%%%%
%% The "tocentry" environment can be used to create an entry for the
%% graphical table of contents. It is given here as some journals
%% require that it is printed as part of the abstract page. It will
%% be automatically moved as appropriate.
%%%%%%%%%%%%%%%%%%%%%%%%%%%%%%%%%%%%%%%%%%%%%%%%%%%%%%%%%%%%%%%%%%%%%

%%%%%%%%%%%%%%%%%%%%%%%%%%%%%%%%%%%%%%%%%%%%%%%%%%%%%%%%%%%%%%%%%%%%%
%% Start the main part of the manuscript here.
%%%%%%%%%%%%%%%%%%%%%%%%%%%%%%%%%%%%%%%%%%%%%%%%%%%%%%%%%%%%%%%%%%%%%
%Introduction
\noindent
Optical spectroscopy plays a pivotal role in studying organic light-emitting materials, which are central to a range of applications in optoelectronics\cite{burroughes1990light}, photonics\cite{samuel2007organic} and bioimaging \cite{giepmans2006fluorescent}, enabling better understanding of their fundamental photophysical properties. However, these materials often exhibit complex excited-state behaviour, and although steady-state absorption and emission spectroscopic techniques can provide useful information, this can only offer a static representation of inherently dynamic processes that occur\cite{bakulin2016ultrafast}. Optical characterisation of these materials is often challenging due to scattering, background fluorescence, and photobleaching. As such, time-resolved spectroscopy can provide this missing dynamical information. In doing so, it reveals discrete processes taking place that govern material performance, e.g., the efficiency of organic semiconductors in an optoelectronic device architecture\cite{long2017effect,shaw2010exciton,nguyen2000controlling,murawski2013efficiency,hedley2017light}. For example, this can allow us to directly observe processes such as exciton generation, exciton diffusion, and charge recombination dynamics on the femtosecond to nanosecond timescale\cite{hayes1997ultrafast}. Therefore, to better understand organic light-emitting materials where fast dynamics can dominate the performance of the material, our primary motivation lies in developing time-resolved spectroscopic techniques that have new advantages. 
\newline
\newline
Conventional spectroscopic techniques for measuring the photoluminescence (PL) lifetime, such as time-correlated single-photon counting (TCSPC), are constrained by photon pile-up effects and shot-noise limitations, especially when probing bulk or highly emissive samples. To address these limitations, recently proposed quantum-enhanced techniques leveraging entangled photon pairs and heralded detection have gained attention as providing a promising alternative solution\cite{li2023single,Harper2023,Eshun2023,Gbler2025,Tsao2025,lvarez-Mendoza2025}. By correlating the arrival times of entangled photons, it is possible to suppress uncorrelated background signals, reducing the effective excitation intensity, and accessing sub-shot-noise sensitivity\cite{Matsuzaki2022}.
As such, this low photon flux is comparable to conditions found in nature, such as natural sunlight\cite{li2023single,Li2025,Fujihashi2020,lvarez-Mendoza2025}. This approach not only offers the potential to investigate bulk organic materials as an alternative to conventional methods, but also opens routes to investigate material optical properties in low fluence regimes. For example, it has recently been theoretically suggested that two-colour photon-correlation measurements can reveal coherent vibronic dynamics during electronic excitation transport when the system is driven under sunlight conditions\cite{Nation2024}. Therefore, utilising low photon flux could emulate the natural sunlight conditions applied to organic solar cells for energy production\cite{nguyen2000controlling,shaw2010exciton}, photosynthesis\cite{collini2010coherently,engel2007evidence}, or phototherapy in healthcare applications\cite{de2016proposed}.
\newline
\newline
Here, we demonstrate quantum-enhanced optical probing of a bulk organic semiconductor poly[2-methoxy-5-(2-ethylhexyloxy)-1,4-phenylenevinylene] (MEH-PPV), an organic dye Rhodamine 6G (Rh 6G), and a bulk fluorescent protein mOrange. We use a superconducting nanowire single-photon detector (SNSPD) and single-photon avalanche diode (SPAD) coincidence detection system for near-infrared and visible correlated photon pairs based on previously proposed theoretical spectroscopic techniques\cite{Fujihashi2024,fujihashi2026two}. Entangled photon pairs are generated via type-0 spontaneous parametric down-conversion (SPDC), with the idler heralding the signal photon that excites the sample. The delay between the heralding detection and subsequent fluorescence emission gives the exciton lifetime, built up over many repetitions to construct a PL decay curve. This approach enables single-photon excitation at visible wavelengths; the nondegenerate nature of the SPDC process further allows the signal to be tuned to shorter wavelengths, extending excitation access to higher-energy transitions. This expands the range of addressable materials beyond what is achievable with conventional excitation sources or degenerate photon pairs. We first verified that the signal photons form a single-photon state, confirming that the fluorophores are indeed excited by single photons. A Ti:Sapphire laser (Coherent Mira900) is frequency-doubled via a lithium triborate (LBO) crystal to produce a 386.6 nm pump, which is focused onto a periodically poled MgO-doped stoichiometric lithium tantalate (PPMgSLT) crystal, generating entangled photon pairs via type-0 SPDC. Periodic poling enables quasi-phase-matching and allows for efficient nonlinear conversion at room temperature. The resulting signal and idler photons are separated and routed to their respective detectors.

\begin{figure}
    \centering
    \includegraphics[width=1\linewidth]{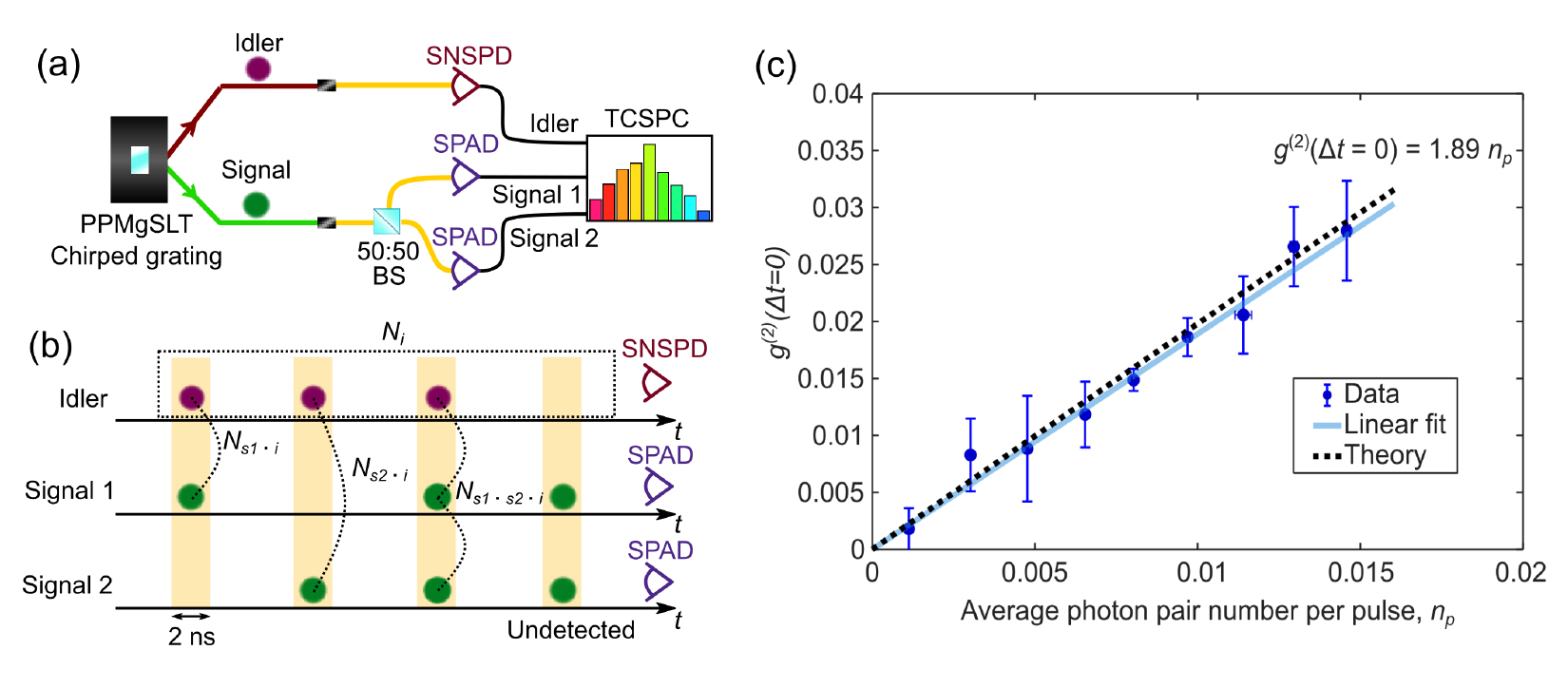}
\caption{Single-photon state confirmation with the three-fold coincidence measurement. \textbf{(a)} Coincidence measurement schematic \textbf{(b)} Time-tagging illustration of coincidence detection between idler and signal photons split by the 50:50 beamsplitter. \textbf{(c)} $g^{(2)}(\Delta t=0)$ plot as a function of average photon pair number per pulse.}
    \label{fig:Heralded_g(2)}
\end{figure}

\noindent
Figure \ref{fig:Heralded_g(2)} (a) shows the experimental schematic (with a more detail description provided in the Supporting Information), where the linearly chirped PPMgSLT crystal is utilised to generate photon pairs. The idler photon, denoted as $i$, is coupled into a single-mode fibre and detected with an SNSPD, whereas the signal photon stream is split into two distinct paths via a 50:50 beam splitter (BS), then captured by SPADs, denoted as $s1$ or $s2$. The detected signals are processed by the TCSPC device. Figure \ref{fig:Heralded_g(2)}(b) depicts the time-tagging picture, in which idler count ($N_{i}$), two-fold coincidence ($N_{s1\cdot i}$ and $N_{s2
\cdot i}$) and three-fold coincidence counts ($N_{s1\cdot s2\cdot i}$) are registered. The coincidence window is set to 2 nanoseconds, triggered by idler count detection. The coincidental second-order coherence $g^{(2)}(\Delta t=0)$ is calculated as\cite{Grangier1986}
\begin{equation}
g^{(2)}(\Delta t=0)=\frac{N_{s1\cdot s2\cdot i}\cdot N_{i}}{N_{s1\cdot i}\cdot N_{s2\cdot i}},
\end{equation}
which is evaluated against the average photon pair number per pulse, $n_{p}$. This value can be calculated as $n_{p}=R_{p}/f_{rep}$, where $R_{p}$ is the photon pair production rate and $f_{rep}$ is the pump laser repetition rate of 76 MHz. The production rate can be calculated from signal and idler count rate ($N_{i}$ and $N_{s}$), and coincidence count rate between them ($N_{s\cdot i}$).  Figure \ref{fig:Heralded_g(2)} (c) shows $g^{(2)}(\Delta t=0)$ as a function of $n_{p}$ along with the linear function fitting. The linear fitting is in good agreement with the theoretical function, and the second-order correlation is well below 0.5, which is the upper boundary of a single-photon state\cite{Grangier1986}. A detailed calculation of the count rate is shown in the Supporting Information.
\newline
\newline
Past reports on heralding emission spectroscopy achieve wavelength tunability through temperature tuning of the nonlinear crystal; adjusting the spectrum to match with the fluorophore absorption band\cite{Harper2023,lvarez-Mendoza2025}. However, we propose a different approach by utilising a linearly chirped grating crystal instead\cite{Harris2007,Nasr2008}. In a quantum-optical experimental context, such entangled photon pairs have been primarily used for quantum infrared spectroscopy, emitting photon pairs in the visible and mid-infrared range\cite{Kalashnikov2016, tashima2024ultra}. 
\newline
\newline
The underlying mechanism of how heralded measurements work is described as follows. A nondegenerate entangled photon pair are generated through type-0 SPDC, excited by a pulsed laser. The photon pair (the signal in the visible and idler in the near-infrared) are frequency-correlated and separated by a dichroic mirror. Correlations between the signal and idler photons cause the detection of one of the photons in each pair to herald its partner. The idler photon is transmitted by the dichroic mirror and simultaneously heralds the signal photon reflected and directed into the sample. The detection of the idler photon acts as the ‘start’ signal for the timing electronics. The sample is excited by the signal photon, emitting a fluorescence photon which is then detected after some time, triggering the ‘stop’ signal. The time difference between the idler ‘start’ (and, by extension, the specific time relative to the idler detection at which the signal photon is incident on the sample), and the fluorescence ‘stop’ is related to how long the exciton exists for. This is repeated to construct a PL lifetime decay. This methodology, as previously reported\cite{Harper2023,Eshun2023,Gbler2025,Tsao2025}, is also achievable with a continuous-wave laser via time-correlated photon counting\cite{fujihashi2026two_JCP}. This enables the visible single-photon excitation necessary for fluorescence lifetime measurements.

\begin{figure}
    \centering
    \includegraphics[width=1\linewidth]{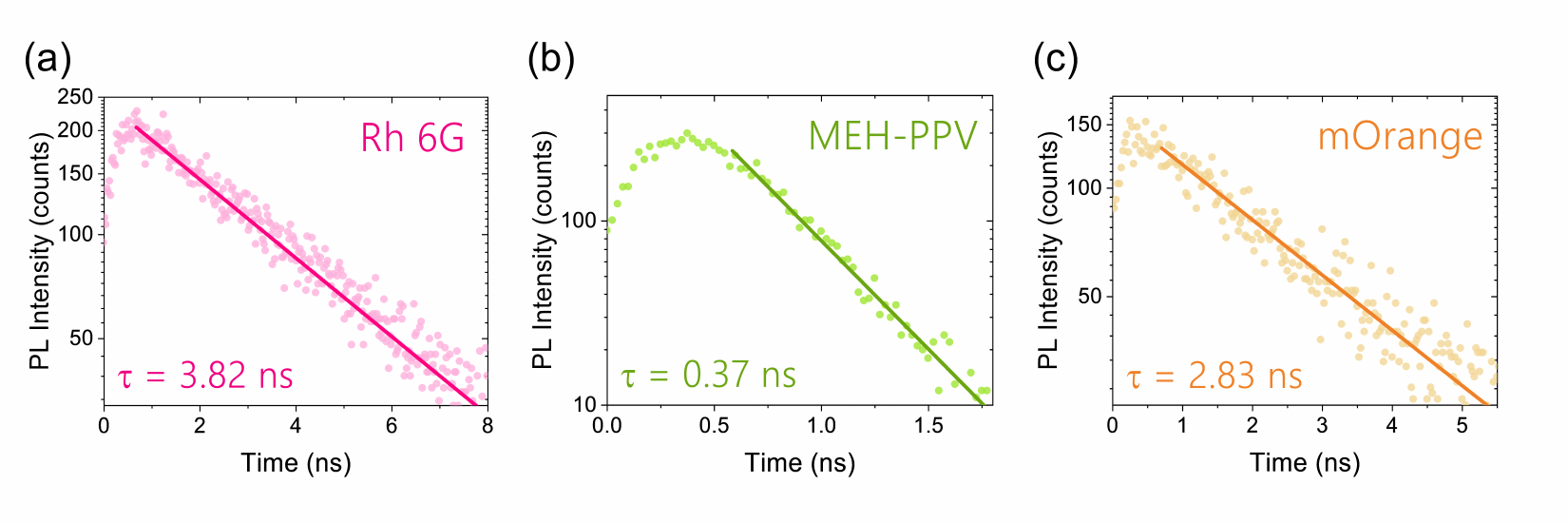}
    \caption{Fluorescence lifetime acquired over 3600 s using the heralded detection scheme from one data set each. PL decays of \textbf{(a)} Rh 6G with a fitted lifetime of 3.82 ns (pink solid line), \textbf{(b)} MEH-PPV with a fitted lifetime of 0.37 ns (green solid line), \textbf{(c)} mOrange with a fitted lifetime of 2.83 ns (orange solid line). The y-axis is logarithmically scaled to give straight lines for exponential decays.} %Rh6G Dataset 1 3600s EP, MEH-PPV Dataset 3 3600s EP, mOrange Dataset 2 3600s EP
    \label{fig:3Lifetimes3600s}
\end{figure}
\begin{figure}
    \centering
    \includegraphics[width=1\linewidth]{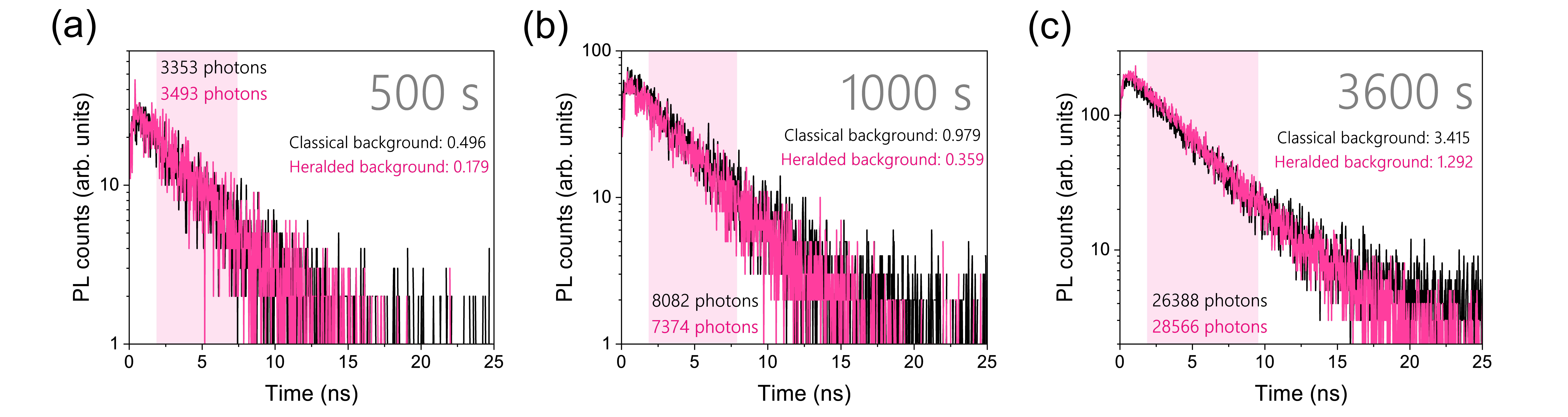}
    \caption{PL decay traces for Rh 6G under classical pulsed-laser excitation (black) and heralded detection (pink) at (a) 500 s, (b) 1000 s and (c) 3600 s integration times. The shaded region (pink) represents the fit range window used for lifetime fitting. The total photon counts within the fit range and mean background photon counts (calculated as the average count per bin over a fixed noise window) is annotated in each panel for the heralded (pink) and classical (black) traces.}
    \label{fig:Rhlifetimes-photon-match}
\end{figure}

\noindent
Using the signal photon from the entangled photon pair SPDC excitation scheme to excite our samples, we first measured PL decay traces for Rh 6G, MEH-PPV, and mOrange, each acquired over 100 s, 500 s and 3600 s. Figure \ref{fig:3Lifetimes3600s} shows mono-exponential fits (solid lines) yielding lifetimes of $\tau$ = 3.82 ns for Rh 6G, $ \tau $ = 0.37 ns for MEH-PPV, and $ \tau $ = 2.83 ns for mOrange at 3600 s. Fluorescence lifetime measurements using (i) heralded single-photon excitation via SPDC will from hereon be referred to as `heralded detection', and (ii) conventional pulsed-laser excitation with TCSPC detection will be denoted `classical pulsed-laser excitation'. The measured Rh 6G lifetime measured is consistent with conventional TCSPC literature values for Rh 6G in water and ethanol\cite{magde2002fluorescence,hanley2001fluorescence,beaumont1993photophysical}, as well as recent SPDC-based entangled photon lifetime measurements reported by Eshun \textit{et al.}\cite{Eshun2023}. This report demonstrated that entangled photon pair correlations can resolve the nanosecond fluorescence decay of Rh 6G with continuous-wave excitation, yielding lifetimes in the $\sim$3.8 - 4.1 ns range across varying concentrations of ethanol and differing objective numerical apertures, remaining in agreement with classical measurements. The sub-nanosecond lifetime obtained for MEH-PPV reflects the fast decays possible with conjugated polymers, coinciding with previously reported conventional TCSPC lifetimes for MEH-PPV in toluene solution\cite{10.1063/1.464790, 10.1063/1.478288, SAMUEL1997497}. Lifetime measurements of mOrange remain comparatively scarce; however, a recent survey reports that fluorescent protein variants, including mOrange measured using classical excitation, exhibit lifetimes spanning $\sim$1.7 - 3.6 ns in solution, with central values around $\sim$2.7 - 3 ns\cite{tan2025time}. 
\newline
\newline
It is instructive to compare these heralded results with their classical equivalents. To highlight the comparison from differences in total photon budget, the classical pulsed-laser excitation power was attenuated so that photon counts within the fit range matched those obtained from heralded detection at each integration time (Fig. \ref{fig:Rhlifetimes-photon-match}). Without matching these fit range counts, differences in signal between the two methods could be attributed to unequal photon counts rather than the detection scheme itself, as seen when the classical excitation power is increased and the classical detection scheme regains its advantage (see Supporting Information).\newline

\noindent
Fig. \ref{fig:Rhlifetimes-photon-match} shows a clear signature of coincidence gating; under matched photon counts, the heralded detection scheme shows a background level approximately 2.6 - 2.8 times lower than the classical pulsed-laser excitation scheme consistently across all three integration times. Heralded counts require temporal correlation with a comparatively rare idler detection event, whereas classical counts are stored whenever the detector registers an event, regardless of whether that photon originated from PL or uncorrelated background. To quantify the uncertainty in each fitted lifetime, we used support plane analysis (SPA), according to an approach described by Lakowicz\cite{Lakowicz2006}, which maps the goodness of fit ($\chi^2_R$) as the lifetime is varied while all other fit parameters are re-optimised. The relative uncertainty is defined as half the width of this confidence interval divided by the fitted lifetime.

\begin{figure}
    \centering
    \includegraphics[width=\linewidth]{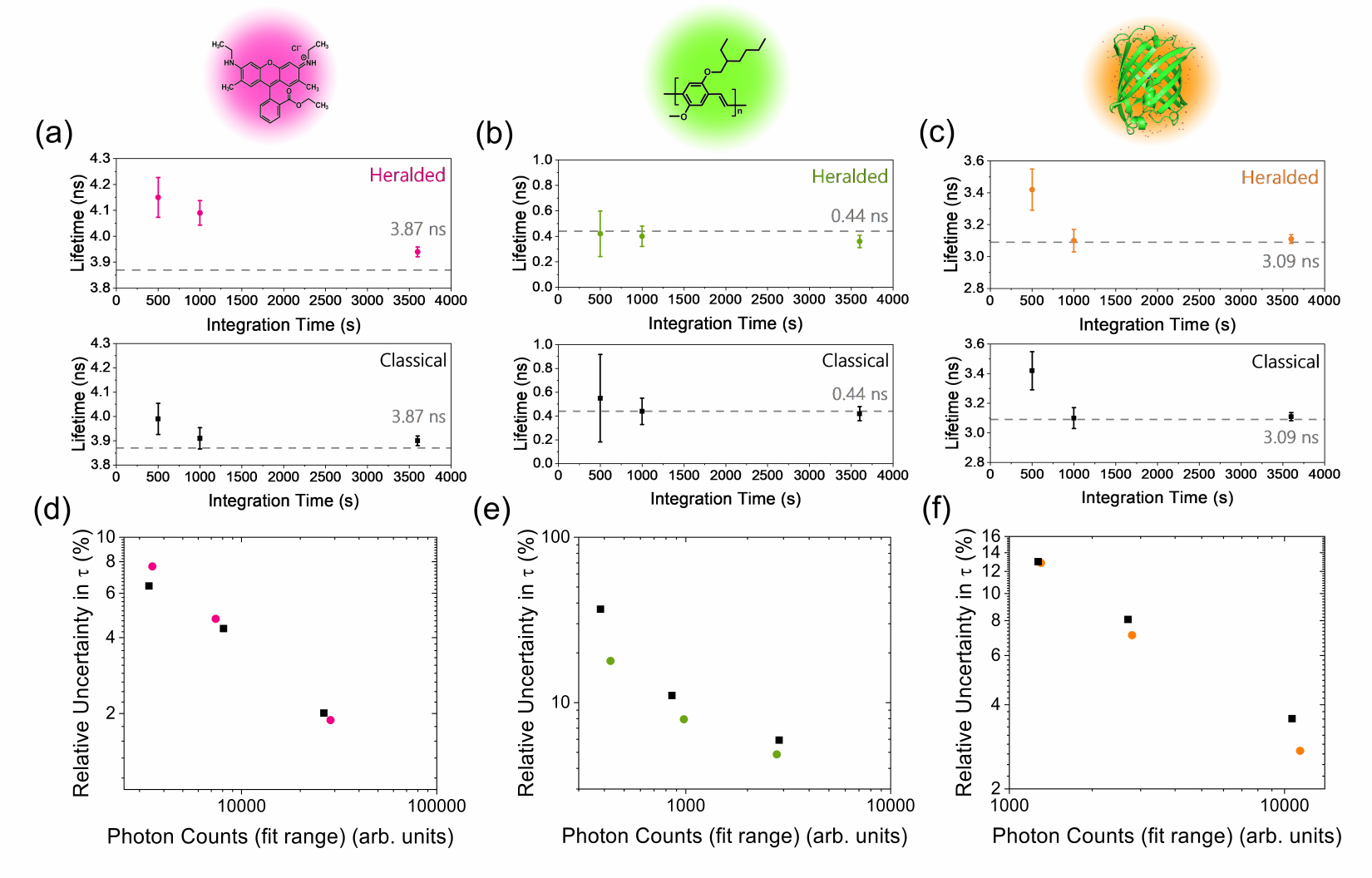}
    \caption{Rh 6G, MEH-PPV molecular structures and mOrange crystal structure (top, left to right) alongside the fitted PL lifetime as a function of integration time for \textbf{(a)} Rh 6G, \textbf{(b)} MEH-PPV and \textbf{(c)} mOrange, each comparing the heralded detection (upper, coloured circle) and classical pulsed-laser excitation (lower, black square) schemes. A reference lifetime obtained from an independent, long-integration classical measurement is included (grey dashed line) as 3.87 ns, 0.44 ns, and 3.09 ns, respectively. Error bars represent asymmetric confidence bounds obtained using SPA. The relative uncertainty in the fitted lifetime (\%) for each material, derived from SPA, as a function of the photon counts within the fit range for heralded (coloured circle) and classical (black square) is shown for \textbf{(d)} Rh 6G, \textbf{(e)} MEH-PPV, \textbf{(f)} mOrange.}
    \label{fig:Lifetime-RelUncert-3materials}
\end{figure}
\begin{figure}
    \centering
    \includegraphics[width=0.75\linewidth]{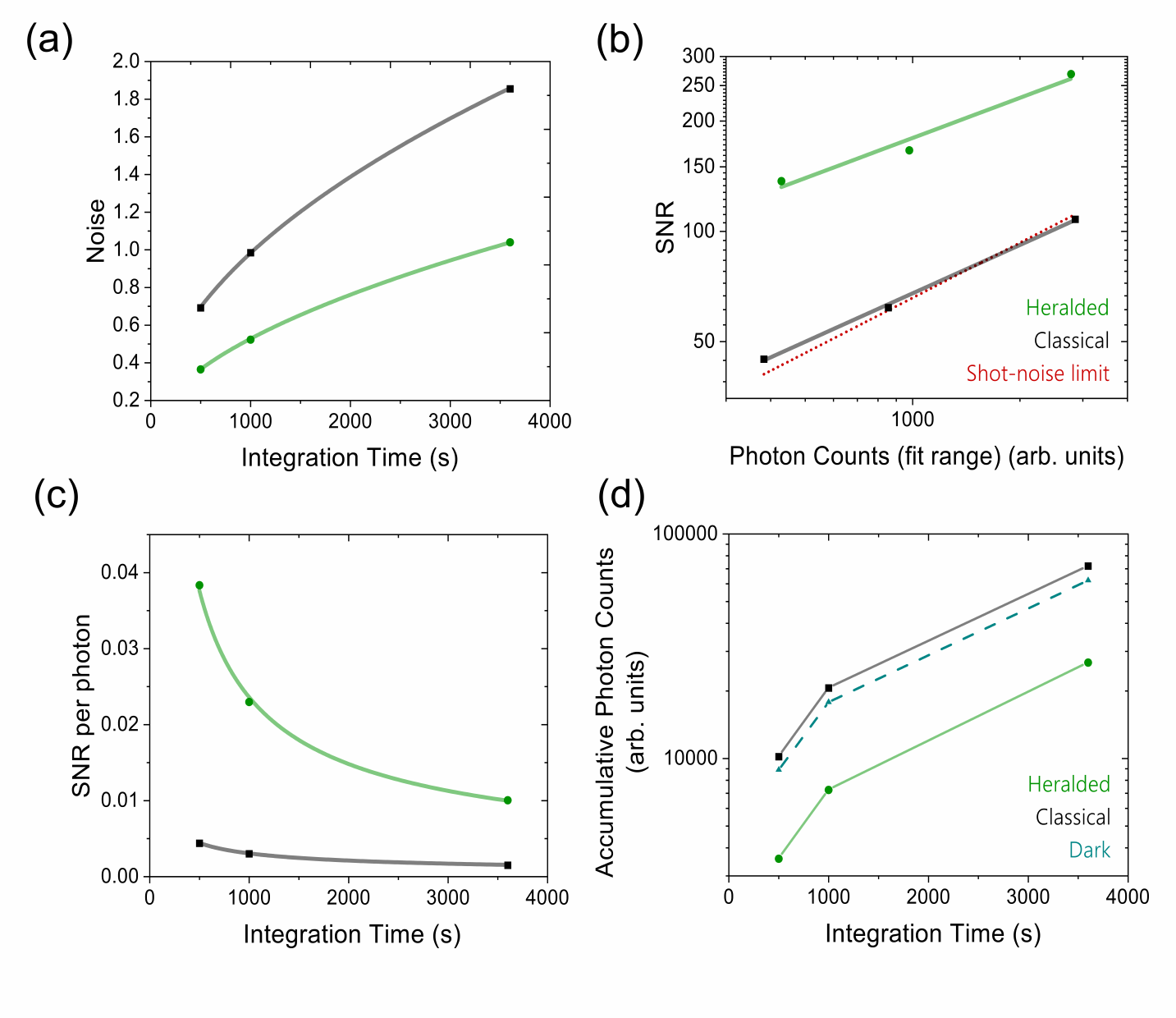}
    \caption{Analysis of the heralded detection advantage in MEH-PPV. \textbf{(a)} Noise over time for heralded detection (green, circles; $N_{\text{heralded}}= 0.013t^{0.53}$) and classical (black, squares; $N_{\text{classical}}= 0.031t^{0.50}$) pulsed-laser excitation. \textbf{(b)} SNR as a function of fit range photon counts for both schemes, plotted alongside a shot-noise-limited reference (red dashed line), $SNR \propto N^{0.5}$. \textbf{(c)} SNR per detected photon plotted against the integration time for both schemes. \textbf{(d)} Total accumulated photon counts across the full measurement window for heralded, classical and dark counts (blue dashed line) against integration time.}
    \label{fig:SNR_Indepe_MEHPPV-SNR-Time}
\end{figure}

\noindent
Fig. \ref{fig:Lifetime-RelUncert-3materials}(a, b, c) extends the comparison of Fig. \ref{fig:Rhlifetimes-photon-match} across the three materials, exhibiting the fitted lifetime across the integration times and testing whether the heralded advantage extends beyond a single material. All materials show that measured lifetimes using both methods converge toward the independently measured reference lifetime with increasing integration time, implying that photostability is maintained for integration times up to 3600 s and that the heralded scheme is sensitive to the same excited-state dynamics. Unlike the other two materials which do not show reliable lifetimes at 500 s for either technique, the heralded detection scheme for MEH-PPV converges more rapidly than for classical with increasing integration times (Heralded 500 s: 0.42 ns, and classical 500 s: 0.55 ns compared to the reference lifetime value of 0.44 ns). This is notable given the well-established heterogeneity of MEH-PPV chains which typically gives rise to variation in measured lifetime  \cite{nguyen2000controlling,shaw2010exciton,schwartz2003conjugated,bout1997discrete}. 
\newline
\newline
This precision advantage of heralded detection is quantified as the relative uncertainty in the fitted lifetime shown in Fig. \ref{fig:Lifetime-RelUncert-3materials}(d, e, f) and appears to be material-dependent in magnitude. For the case of MEH-PPV (Fig. \ref{fig:Lifetime-RelUncert-3materials}e), heralded detection achieves a much smaller relative uncertainty than classical detection at every integration time and showcases the heralded advantage consistently. However, for Rh 6G (Fig. \ref{fig:Lifetime-RelUncert-3materials}d), where the photon budget over the fit range for heralded detection and classical laser-pulsed excitation schemes are most closely matched, the heralded scheme only slightly out-competes the classical scheme on the measurement with the highest number of photon counts within the fit range. For mOrange (Fig. \ref{fig:Lifetime-RelUncert-3materials}f), the relative uncertainties for both schemes are comparable at the lowest photon count within the fit range, with the heralded relative uncertainty decreasing with increasing photon counts. The heralded advantage therefore becomes increasingly apparent with increasing photon counts for this particular material. 
\newline
\newline
In a previous report by G\"abler \textit{et al.}\cite{Gbler2025}, the signal-to-noise ratio (SNR) of the instrumental response function and that of the heralded emission were compared. It was found that the SNR of the IRF is higher than that of the heralded emission due to the low brightness of the heralded emission. However, the impact on the SNR when using an entangled photon rather than a classical laser have not yet been explored. To address this, we compare between heralded detection and classical pulsed-laser excitation schemes and discuss the integration time dependence of SNR. The SNR is defined as,  
\begin{equation}
    SNR_{\tilde{F}} = \frac{\tilde{F}_{\text{max}}}{\sigma_{\tilde{F}}},
    \label{eq:SNR}
\end{equation}
where $\tilde{F}_{\text{max}}$ and $\sigma_{\tilde{F}}$ represent the maximum counts of the fluorescence peak and the standard deviation, respectively\cite{gili2022experimental}. The standard deviation is calculated using the background counts measured early in the time window, which is expressed as \begin{equation}
    \sigma_{\tilde{F}} = \sqrt{\frac{1}{N_{t}}\cdot\sum_{i=0}^{N_{t}} \tilde{F}^{2}(t_{i})},
\end{equation}
where the $N_{t}$ is the number of sample bins, which in this case is 300, and $t_{i}$ is the index of the summed bins.
\newline
\newline
To deepen our understanding of the origin of the consistent heralded advantage observed specifically with MEH-PPV in (Fig. \ref{fig:Lifetime-RelUncert-3materials}b,e), we explore the SNR relationship in Fig. \ref{fig:SNR_Indepe_MEHPPV-SNR-Time}. For both methods, the measured noise follows a square-root dependence on integration time as shown in Fig. \ref{fig:SNR_Indepe_MEHPPV-SNR-Time} (a), which is consistent with ordinary photon-counting statistics as indicated in the caption by the exponents of 0.53 and 0.50 for heralded detection and classical pulsed-laser excitation, respectively. The noise floor for the classical scheme is higher than that of the heralded detection throughout. The heralded detection methodology maintains a clear advantage over a fitted classical shot-noise reference which becomes apparent when we consider how SNR relates photon counts within the fit range (Fig. \ref{fig:SNR_Indepe_MEHPPV-SNR-Time}(b), $SNR_{\text{heralded}}=15.79t^{0.34}$ and $SNR_{\text{classical}}= 2.87t^{0.44}$.  \newpage
\noindent
The SNR for heralded detection is approximately 2.4 to 3.0 times higher than that of classical pulsed-laser excitation (when directly calculated from the fitted equation) across the increasing photon counts in the fit range, though this gap narrows slightly as the counted number of photons increase. This is consistent with the two methods showing slightly different scaling behaviour with photon number (b $\approx $ 0.34 for heralded, b $\approx $ 0.44 for classical, against a shot-noise-limited b = 0.5). This pattern also occurs for mOrange and Rh 6G, with both the noise and SNR comparisons showing the same behaviours for both materials. For Rh 6G, it's heralded and classical noise exponents are 0.53 and 0.49, respectively, and for mOrange the noise exponents for both heralded and classical are 0.52 and 0.49. All noise exponents fall within the standard shot-noise reference value of 0.5, and the heralded detection's noise exponent consistently exceeds the classical excitation's noise exponent values by a small margin across all three materials independently. As for the SNR of Rh 6G ($SNR_{\text{heralded}}= 15.51t^{0.31}$ and $SNR_{\text{classical}}= 4.07t^{0.41}$) and mOrange ($SNR_{\text{heralded}}= 6.47t^{0.35}$ and $SNR_{\text{classical}}= 2.61t^{0.39}$), the same pattern emerges; the heralded advantage over classical is approximately 1.7 to 2 times for Rh 6G and 1.8 to 1.9 times for mOrange across the integration times. The power-law parameters extracted from the mOrange and Rh 6G SNR fits carry a larger relative uncertainty than for MEH-PPV and are reported to further support the general trend (see Supporting Information).
\newline
\newline
By normalising the SNR by the total number of photons detected over the measurement window (Fig. \ref{fig:SNR_Indepe_MEHPPV-SNR-Time}c) indicates that heralded detection extracts approximately 6.7 - 8.7 times more SNR per detected photon than classical excitation for MEH-PPV across the integration times studied (0.0383 vs. 0.0044 at 500 s; 0.0101 vs. 0.0015 at 3600 s). This advantage is higher than that observed in raw SNR as the heralded detection also registers roughly 35 to 37\% as many total photons, with the uncorrelated background being rejected at detection. Figure \ref{fig:SNR_Indepe_MEHPPV-SNR-Time} (d) allows us to identify the source of the heralded advantage. The heralded detection method's accumulative photon counts across the entire measurement window are consistently lower than that of the classical excitation scheme. For each integration time, the total number of heralded photon counts across the entire measurement window is approximately a third of the total number of photon counts for the classical excitation scheme (heralded = 3575, classical = 10186 at 500 s, heralded = 7240, classical = 20625 at 1000 s, and heralded = 26729, classical = 72022 at 3600 s), despite the two method's photon counts being matched within the fit range itself. 
\newline
\newline
This occurs because the classical excitation scheme's accumulative photon counts over the entire measurement window is diluted by background counts that lie outside the fit range; approximately 3.7 to 4.1\% of the accumulative total counts reside within the fit range used for the lifetime fit, with the remaining photons contributing to background across the rest of the measurement window. However, through background suppression, heralded detection retains 10.4 to 13.5\% of its total within this region, which is 2.6 to 3.2 times that of the classical schematic at every integration time. The classical schematic's total photon counts significantly exceeds the dark-count level (8894, 17755, and 62195 photons at 500 s, 1000 s, and 3600 s, respectively) at every integration time. Using this, we can confirm that real PL signal is present for both techniques, however, the heralded detection's accumulative photon counts sit well below the dark-count level throughout, reflecting the technique's deliberate exclusion of background signal. The heralded advantage stems from a substantially lower absolute noise floor under heralded detection, which is a direct consequence of coincidence gating suppressing uncorrelated background counts that classical detection cannot otherwise avoid accumulating. Interestingly, the SNR advantage does not arise from the heralded scheme's noise growing more slowly - this increases with integration time at comparable rates under both schemes. It presumably originates from arises from a lower noise amplitude at every integration time with a smaller contribution from a higher peak signal and is most pronounced for MEH-PPV where both are largest.
\newline
\newline
The observed effect is important for materials whose photophysical behaviour is strongly influenced by excitation intensity. In the case of conjugated polymers, these possess $\pi$-$\pi$* transitions which typically lie within the visible range; therefore, excitation within this absorption band efficiently populates singlet excitons. Furthermore, by avoiding high intensity illumination, excitation within this band limits the formation of high exciton densities and associated nonlinear excitonic processes, thereby reducing their influence on exciton diffusion and trapping behaviour \cite{nguyen2000controlling, shaw2010exciton,Fairbairn2025}. Processes such as these can act as loss mechanisms, reducing the overall light-emitting efficiency of the material. Moreover, single-photon excitation enables fluorescence lifetime measurements of materials and systems (for example, thin film devices, narrow-band organic emitters or low-bandgap polymers) where conventional high-intensity excitation may modify their morphology or degrade them altogether\cite{scholz2015degradation, aziz1999degradation,reineke2013white,baldo1999excitonic,jorgensen2008stability,manceau2011photochemical}. For fluorescent proteins such as mOrange, these systems are engineered to operate within the visible region to minimise photodamage and potential for undesirable photochemical side-reactions. Using a heralded detection scheme enables access to the emissive singlet state at low intensities, aligned with developments in emerging quantum-enhanced spectroscopic approaches in biological chromophores\cite{scholes2011lessons,kim2019venusa206}. Similarly, for Rh 6G, low intensity visible excitation suppresses excited-state absorption and photobleaching, allowing intrinsic fluorescence lifetimes to be observed free from intensity-induced photothermal effects and associated artefacts\cite{widengren1996mechanisms,eggeling1998photobleaching,zhang2023characterizations, zipfel2003live, Denk1990}, rendering it suitable for fluorescence lifetime measurements using the heralded detection scheme\cite{Eshun2023}. 
\newline
\newline
The noise suppression observed in heralded detection reflects a fundamental difference in the photon statistics of the two excitation schemes. Classical pulsed-laser excitation follows Poissonian photon statistics, while heralded detection approximates a Fock state. In the heralded scheme, the coincidence gate filters only those detection events correlated with a herald photon; this is a mechanism that has no analogue in classical excitation. This distinction is captured by comparing the total detection efficiencies of the two schemes. For the heralded case, the signal detection efficiency is given by $\eta_{s}$= $N_{s\cdot i}$/$N_{ i}$, and the idler detection efficiency by $\eta_{i}$= $N_{s\cdot i}$/$N_{ s}^{*}$, where $N_{s\cdot i}$ are the coincidence counts, $N_{ i}$ are the idler counts, and $N_{ s}^{*}$ represents the signal counts with the sample present. For classical, the analogous quantity is $\eta_{\text{classical}}$ = $N_{1}$/$N_{ p}$, where $N_{1}$ are the detected fluorescence counts and $N_{p}$ are the incident pump photons. The ratio $\eta_{i}$/$\eta_{\text{classical}}$ therefore provides a quantitative measure of the enhancement, while the complementary ratio (1 - $\eta_{i}$)/(1 - $\eta_{\text{classical}}$) captures the relative suppression of undetected events. Separating both signal and noise contributions of MEH-PPV, yields amplitude prefactors of 0.0134 and 0.0313 for heralded and classical excitation respectively, giving a noise suppression ratio of approximately 0.43. The detection efficiency ratios alone are insufficient to fully account for this suppression; a comprehensive analysis instead requires a quantum statistical description that considers the non-Poissonian nature of the heralded photon pairs. The consistent nature of this enhancement across all material classes indicates that optical losses - rather than any material-specific property - are the dominant factor governing the degree of improvement over classical excitation.
\newline
\newline
The advantages demonstrated here are relevant to fast fluorescence lifetime imaging microscopy (FLIM), live-cell imaging and single-molecule spectroscopy, especially where low photon flux and rapid acquisition are often essential. More broadly this is particularly suitable for photophysically fragile soft-matter and biological systems\cite{liu2019fast}. The three material classes studied can be applied across optoelectronics, bioimaging and molecular sensing \cite{Lakowicz2006}. Utilising nondegenerate SPDC opens up pathways towards heralded spectroscopic and imaging modalities that can exploit photon-pair correlations. This can then be exploited to access additional spatial, spectral and dynamic information at similarly low flux \cite{dorfman2016nonlinear}, including heralded FLIM, heralded excitation spectroscopy (spectral information from the idler photon alone) and heralded fluorescence correlation spectroscopy (FCS)\cite{magde1972thermodynamic}. Heralded FCS is particularly compelling because most FCS applications involve biological systems, including fluorescent proteins such as mOrange, where low excitation is crucial to minimise photobleaching and phototoxicity\cite{haustein2003ultrasensitive,shaner2005guide}. In summary, we have demonstrated PL lifetime measurements using heralded detection of nondegenerate photons pairs using SPDC and compared this against classical pulsed-laser excitation for Rh 6G, MEH-PPV and mOrange. At matched photon counts within a defined fit range, the heralded detection technique yields an SNR which is approximately 1.6 - 3.0 times higher than classical excitation across all three materials, with consistent lifetimes being achieved at long integration times. Measured lifetimes of MEH-PPV using the heralded schematic at short integration times of 500 s outperforms classical excitation. The quantum advantage arises from coincidence gating suppressing uncorrelated background, which reduces the noise amplitude of MEH-PPV to less than half of that of classical excitation; with the main limitation in the performance of the technique originating from the detector saturation or SPDC generation limit. Heralded spectroscopy decouples detected signal quality from photon flux, whilst classical laser spectroscopy's noise is governed by flux-dependent effects, rather than total photon count of integration time. 

\section{Experimental Methods}\label{sec5}

\subsection{Heralded emission spectroscopy}
The pump source used is a mode-locked Ti:S laser, with a peak at approximately 773.2 nm, a spectral linewidth of 0.2 nm and a repetition rate of 76 MHz. This excitation light is used in conjunction with a 12 mm-long LBO crystal to generate second-harmonic light. This is then normally incident on a PPMgSLT crystal at a power ranging from 5 to 85 mW, generating nondegenerate entangled photon pairs through SPDC. The entangled photon pair are separated using a dichroic mirror, which reflects the signal photons ($\sim$515 nm) and transmits the idler photons ($\sim$1550 nm). The idler photon is simply coupled into a single-mode fibre and detected by the SNSPD. The reflected signal photons are coupled using a single-mode fibre, directed into a separate microscope objective (NA = 0.66) set-up used to excite the fluorescence samples. The emitted photons are then collected using a multi-mode fibre and followed by a fibre-coupled SPAD. The time-to-digital converter analyses the photon arrival times from the idler and PL photons and records coincidence counts to construct the PL lifetime decay.  A more comprehensive description of the optical set-up is available in the Supporting Information. 

\subsection{Sample Preparation}
Rh 6G was dissolved in water and mixed thoroughly using a vortex mixer, giving a concentration of 41.37 $\mathrm{\mu M}$. MEH-PPV was dissolved in toluene and mixed, producing a solution with a concentration of 40.0 $\mathrm{\mu M}$. Toluene was chosen due to MEH-PPV being highly soluble in this solvent. The mOrange fluorescent protein was expressed in E. coli BL21(DE3) cells upon IPTG induction. Cell lysates were prepared in 20 mM Tris-HCl buffer (pH 8.0) supplemented with 0.5 M NaCl, and the protein was purified sequentially using a HisTrap HP column followed by a HiTrap Q XL column (Cytiva). The resultant mOrange concentration was 113.2 $\mathrm{\mu M}$. Using a pipette, 100 $\mathrm{\mu L}$ of Rh 6G and mOrange solutions were each measured in $\sim$170 $\mathrm{\mu M}$ thick sterilised borosilicate glass bottom dishes situated above the objective. For MEH-PPV, 600 $\mathrm{\mu L}$ of the solution was measured using a $\sim$170 $\mathrm{\mu M}$ thick sterilised borosilicate glass coverslip located within a home-built air-tight chamber to prevent evaporation throughout the measurement. 

\subsection{Fluorescence Lifetime Analysis}
The fluorescence lifetimes were extracted from the TCSPC decay data by single-exponential tail fitting using a custom Python-based analysis program. Prior to fitting, the decay data was background subtracted. Fitting was performed using a non-linear least squares optimiser with Poisson (shot-noise) weighting, and  uncertainties as well as fit quality were determined using SPA. The $\chi^2_R$ surface is mapped by stepping each fitted lifetime parameter across a grid of fixed values while re-optimising all other free parameters with each step. The 1$\sigma$ confidence interval is defined as the range over which $\chi^2_R$ remains below the threshold \begin{equation}
\chi^2_R(\text{par}) = \chi^2_R(\text{min})\left(1 + \frac{p}{\nu}F(p,\nu,0.68)\right),
\end{equation}
where $p$ represents the number of free parameters, $\nu$ is the degrees of freedom, and $F(p,\nu,0.68)$ is the F-statistic at $P=0.68$. The value of $P=0.68$ corresponds to the $\pm$1$\sigma$ for  a Gaussian-distributed quantity.

\section{Acknowledgements}\label{sec7}
N.J.F. was supported by funding from Japan Society for the Promotion of Science (JSPS) through both a JSPS International Postdoctoral Fellowship (P24767) and KAKENHI Grant No. 25KF0034. We acknowledge funding from the MEXT Quantum Leap Flagship Program (MEXT Q-LEAP) Grant No. JPMXS0118069242. R.S. and G.J.H. acknowledge funding from the JSPS-UKRI International Joint Research Program (JRP-LEAD with UKRI; JSPS Grant No. JPJSJRP20241704 and UKRI Grant No. UKRI659). G.J.H. also acknowledges UKRI funding EP/V048805/1. The SNSPD devices were fabricated using the facilities of the Advanced ICT Device R\&D Promotion Center, NICT.

\section{Supporting Information}\label{sec8}
Additional experimental details including schematic of experimental setup, signal-to-noise ratio analysis for additional materials, absorption and emission spectra, photon fluence plot, table of fit parameters, reference lifetime plots, support plane analysis profiles, detector instrument response functions, characterisation of the heralded excitation source and detection.

\bibliography{achemso-demo}

\end{document}